\documentclass[sn-mathphys,Numbered]{sn-jnl}

\usepackage{graphicx}%
\usepackage{multirow}%
\usepackage{amsmath,amssymb,amsfonts}%
\usepackage{amsthm}%
\usepackage{mathrsfs}%
\usepackage[title]{appendix}%
\usepackage{xcolor}%
\usepackage{textcomp}%
\usepackage{manyfoot}%
\usepackage{booktabs}%
\usepackage{algorithm}%
\usepackage{algorithmicx}%
\usepackage{algpseudocode}%
\usepackage{listings}%
\begin{document}
\title[Beyond mean-field effects in dynamics of  BEC...]{Beyond mean-field effects in dynamics of  BEC in the double-well potential}

\author[1,2]{\fnm{Fatkhulla  Kh.} \sur{Abdullaev}}\email{fatkhulla@yahoo.com}
\author*[1]{\fnm{Ravil M.} \sur{Galimzyanov}} \email{RavilGalimzyanov2024@gmail.com}
\author[1]{\fnm{Akbar M.} \sur{Shermakhmatov}} \email{shermaxmatovakbar@gmail.com}

\affil[1]{\orgdiv{Theoretical physics department}, \orgname{Physical-Technical Institute of the Uzbekistan Academy of Sciences}, \orgaddress{\street{2-B, Ch. Aytmatov}, \city{Tashkent}, \postcode{100084},  \country{Uzbekistan}}}

\affil[2]{\orgname{Institute of Theoretical Physics, National University of Uzbekistan}, \orgaddress{\street{Str. University}, \city{Tashkent}, \postcode{100174},  \country{Uzbekistan}}}

\abstract{
The nonlinear dynamics of a quasi-1D BEC loaded in a double-well potential is studied. The beyond mean-field corrections to the energy in the form of
the Lee-Huang-Yang term are taken into account. One-dimensional geometry
is considered. The problem is described in the scalar approximation by the
extended Gross-Pitaevskii (EGP) equation with the attractive quadratic nonlinearity, due to the Lee-Huang-Yang correction, and the effective cubic mean-field nonlinearity describing the residual intra- and inter-species interactions. To describe tunneling and localization phenomena, a two-mode model was obtained. The frequencies of the Josephson oscillations are found and confirmed by the full numerical simulations of the EGP equation.
The parametric resonance in the Josephson oscillations, when the height of the barrier is periodically modulated, is studied.  The predictions of the dimer model, including the case of the one-dimensional Lee-Huang-Yang superfluid, have been proven.}




\maketitle
\section{Introduction}
\label{intro}
The investigation of the beyond-mean field effects in the BEC dynamics
has attracted a lot of attention in the last years. The beyond mean field
energy corrections to account for the effects of
quantum fluctuations have been calculated by Lee-Huang-Yang  in the work~\cite{LHY}.
These corrections give a small contribution to the  dynamics of a one-component atomic  BEC. The situation is changed drastically in the case of a two-component BEC, when the mean field effects are balanced, so the residual mean-field contribution to energy is
comparable with the Lee-Huang-Yang( LHY) correction. As result  the
unstable against collapse BEC can be stabilized by the quantum
fluctuations~\cite{Petrov,PA}.

In this way, quantum droplets are generated. This prediction has
been confirmed in the experiments~\cite{exp1,exp2}.
The quantum droplets also can exist in one component dipolar BEC,
where the attractive residual mean filed interactions from the
two-body and dipolar interactions are balanced by the LHY quantum
correction~\cite{dip1,dip2,dip3}. Modern status of the area
can be found in the reviews~\cite{rev1,rev2}.
Many other effects of quantum corrections on processes of  collective
oscillations, modulational instability, and the miscible-immiscible
transition in BEC, have been performed recently~\cite{collexc,KL,MI,MIT}.

One interesting is the fundamental problem of quantum tunneling
and localization of BEC in the double-well trap  under quantum
fluctuations, i.e. the role of the  beyond mean field effects for
this problem.
In the mean-field approach, macroscopic quantum tunneling (MQT) and
self-trapping (ST) in a double-well potential were investigated in~\cite{TM}. It was shown that nonlinear interactions play an
essential role in the Josephson oscillations and appearance of
the self-trapped states.

Since the  LHY term in the GP equation have the nonlinear form,
quantum fluctuations (QF) should strongly  influence the MQT and
ST. In a quasi-one-dimensional geometry with $l_{\perp} \gg a_s$ where $l_{\perp}$ is the transverse oscillation, and $a_s$ is the atomic scattering length, this problem is considered in~\cite{AGS}. Two-mode models with the LHY term for (1D-3D) dimensions are considered in~\cite{Salas}. In this case, QF lead to the effective repulsive quartic interaction.

The case of the one-dimensional geometry,  when $l_{\perp} \leq a_s$
corresponds to quantum fluctuations giving an effective attractive
quadratic nonlinearity in the GP equation. Thus crossover from
quasi-one-dimensional geometry to one-dimensional is described by
changing sign of the effective nonlinearity from the repulsive
to attractive one.

In this work, we will study the dynamics of two-component quasi-one-dimensional  BEC loaded into a double-well trap when  quantum fluctuations give the effective attractive quadratic interaction term in the GP equation.

  The structure of the paper is as follows.
In Section~\ref{sec:aver}, we describe a modified
quasi-one-dimensional GP equation, obtained for
the two-component BEC
in a scalar approximation including the LHY term.
In Section~\ref{sec:twomodes}, a two-mode model for a BEC loaded
into a double-well trap is derived. The Hamiltonian, fixed
points, the Josephson oscillations  are analyzed in
Section~\ref{dimer}. The nonlinear self-trapped regime
is also discussed.
 The analytical  results are compared with full numerical
 simulations of the modified Gross-Pitaevskii (GP) equation with the
double-well potential.
 In Section~\ref{sec:resonance}, the dynamics in the trap with
the periodically modulated in  time barrier height is investigated.
In conclusion the obtained results are summarized.

\section{The modified Gross-Pitaevskii equation  for a BEC in a double-well  potential}
\label{sec:aver}
We  consider the dynamics of a two-component BEC loaded in a
cigar-type trap, when the beyond mean-field effects, describing
quantum fluctuations are taken into account.
The system of  modified GP equations has the form~\cite{PA,Mithun}:
\begin{eqnarray}
i\hbar\Psi_{1,T}=-\frac{\hbar^2}{2m}\Psi_{1,XX}  +V(X)\Psi_1 + (g_{11}|\Psi_1|^2 + g_{12}|\Psi_2|^2)\Psi_1 - \frac{g_{11}\sqrt{m}}{\pi \hbar} \nonumber \\
(g_{11}|\Psi_1|^2 + g_{22}|\Psi_2|^2)^{1/2}\Psi_1,\\
i\hbar\Psi_{2,T}=-\frac{\hbar^2}{2m}\Psi_{2,XX}  + V(X)\Psi_2 +(g_{22}|\Psi_2|^2 + g_{21}|\Psi_2|^2)\Psi_2- \frac{g_{22}\sqrt{m}}{\pi\hbar}
\nonumber \\
(g_{11}|\Psi_1|^2 + g_{22}|\Psi_2|^2)^{1/2}\Psi_2,
\end{eqnarray}
where    $g_{ij}= 2\hbar\omega_{\perp}a_{ij} \ $, $\omega_{\perp}$  is
the transverse atomic scattering length,
and  $a_{ij}$ are the  intra- and inter-spieces atomic scattering lengths.
If $g_{11}=g_{22}=g, \ g_{12}=g_{21}$,
   the scalar approximation can  be used
$
\Psi_1=\Psi_2 =\Psi.
$
Then we have the modified GP-equation:
\begin{equation}
i\hbar\Psi_T = -\frac{\hbar^2}{2m}\Psi_{XX} + \delta g |\Psi|^2\Psi -\frac{\sqrt{2m}}{\pi\hbar}g^{3/2}|\Psi|\Psi+
V(X)\Psi.
\end{equation}
Here the double-well potential is:
\begin{equation*}
V(X)=\frac{1}{2}m\omega_x^2 X^2 + V\mbox{sech}^2(-\frac{X}{d}),
\end{equation*}
and
\begin{equation*}
\delta g = g_{12} + \sqrt{g_{11}g_{22}}, g=\sqrt{g_{11}g_{22}}.
\end{equation*}
This equation has recently been used for the investigation of the scattering of one-dimensional quantum droplets by narrow and broad barriers and wells in works~\cite{Hu2023,Malomed2023}.

The interesting case is when $\delta g = 0$ and the residual mean-field contribution is equal to zero, corresponds to the Lee-Huang-Yang fluid~\cite{Jor,Skov}. In this instance, the dynamics of a BEC in the double-well trap is controlled by the effective nonlinearity given by the quantum fluctuations.

It is useful to rewrite the equation in the dimensionless form. Let us
introduce dimensionless variables:
\begin{eqnarray*}
t&=&T\omega_{\perp}, \ x=X/l_{\perp}, \ l_{\perp}=\sqrt{\frac{\hbar}{m\omega_{\perp}}}, \ V_0=\frac{V}{\hbar\omega_{\perp}}, \ l=\frac{d}{l_{\perp}}, \\ \bar{\psi}&=& \frac{(2m)^{1/2}}{\pi\hbar^2\omega_{\perp}}g^{3/2}\psi, \
 \gamma = \frac{\pi^2}{8}\left(\frac{l_{\perp}^2 \delta g}{a^2 g}\right).
\end{eqnarray*}
We have:
\begin{equation}\label{eq4}
i\psi_t =-\frac{1}{2}\psi_{xx} + \gamma |\psi|^2\psi - |\psi|\psi + V(x)\psi,
\end{equation}
where
\begin{equation}
V(x)= \frac{l_{\perp}^2}{2 l_{x}^2}x^2 + V_0 \mbox{sech}^2 \left(-\frac{x}{l}\right).
\end{equation}
The number of atoms $ N=\int dx |\psi|^2$  in the physical units is:
\begin{equation}
N = \frac{16 a^3}{\pi^2 l_{\perp}^2}N_p.
\end{equation}
Note that this equation is valid when the condition $l_{\perp} \sim  a$
is satisfied.

The values of parameters for the experiment are:
$(l_{\perp}/ l_x)^2 \sim 10^{-2}, \ \gamma \gtrsim 0.1, \ V_0 \sim 1, l \geq 0.5$ and $N \sim 10$ for $N_p \sim 10^3, \ a \sim 0.3 l_{\perp}$.

\section{Two-mode model} \label{sec:twomodes}

\begin{figure}[htbp]\label{fig1}
\includegraphics[width=0.9\textwidth]{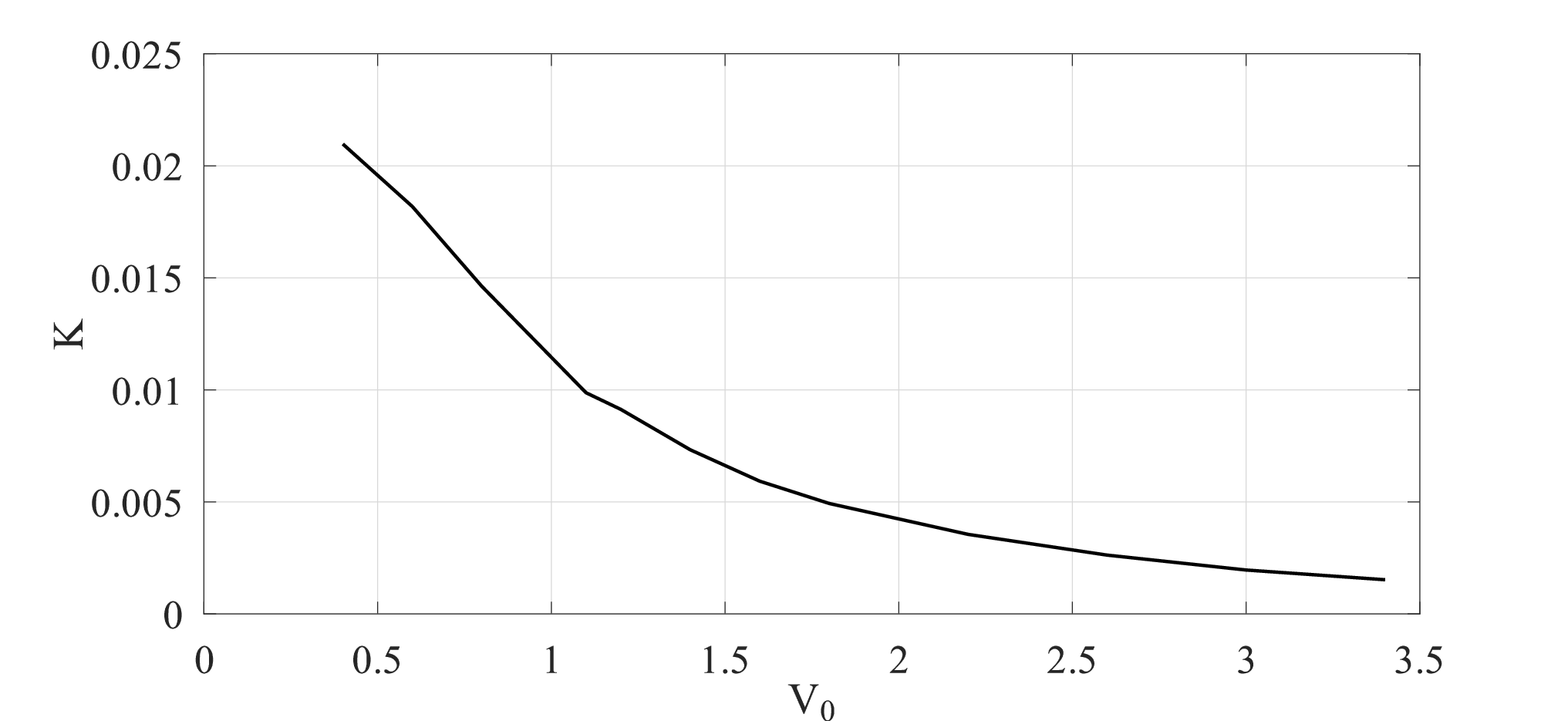}\quad
\caption{The dependence of the amplitude of the tunneling K on the potential barrier $V_0$. Here, $\gamma=0.1$, $l_{\perp}/l_x=0.1$, \ $l=1$  }.
\end{figure}

For investigation of the BEC dynamics in a double-well trap,
it is useful to employ a two-mode description. We will apply the procedure
analogous developed in~\cite{TM}. First, we find wave-functions for the modes,
ground, and first excited states, which can be found from the solution of the
eigenvalue problem with $\psi = \phi \exp(-i\mu t)$, where $\mu $ is the chemical potential:
\begin{equation}
\mu \phi =-\frac{1}{2}\phi_{xx} + \gamma \phi^3 -\phi^2 + V(x)\phi.
\end{equation}
We will look for a solution of the form:
\begin{equation}
\phi = u(t)\phi_1(x) + v(t)\phi_{2}(x),
\end{equation}
where
\begin{equation}
\phi_1 = \frac{\phi_g + \phi_e}{\sqrt{2}}, \ \phi_2 = \frac{\phi_g - \phi_e}{\sqrt{2}}.
\end{equation}
Here $\phi_g, \phi_e $ are eigenfunctions of the ground state and the first
excited state respectively.
They satisfy the orthogonality condition $\int \phi_i\phi_j dx = \delta_{ij}$.
Substituting these expressions into the modified GP equation~(\ref{eq4}),
after the integration $x$ and ignoring  nonlinear overlap integrals, we
obtain equations for the two-mode model:
\begin{equation}\label{dimer_eq1}
iu_t - \bar{\gamma} |u|^2 u + \bar{g} |u|u +Kv + E_1 u=0,
\end{equation}
\begin{equation}\label{dimer_eq2}
iv_t - \bar{\gamma} |v|^2 v + \bar{g} |v|v +Ku + E_2 v=0.
\end{equation}
Here
\begin{eqnarray*}
\bar{\gamma} =\gamma \int dx \phi_1^4, \ \bar{g}= \int dx \phi_1^3, \  K= -\left[\frac{1}{2} \int dx \phi_{1,x} \phi_{2,x} + \int dx \phi_1 V(x)\phi_2 \right],
\end{eqnarray*}
\begin{eqnarray*}
 E_i = \frac{1}{2} \int |\phi_{i,x}|^2 dx + \int \phi_i V(x) \phi_i dx,
\end{eqnarray*}
$\bar{\gamma}$ and $\bar{g}$ are nonlinear parameters, $E_{1,2}$ are zero-point energies in each trap, $K$ is the amplitude of the tunneling. Fig.~\ref{fig1} depicts the dependence of $K$ on the barrier height $V_0$. As seen the tunneling parameter $K$ decreases as the potential barrier height increases. Note that the two-mode model is applicable when the chemical potential is less then the barrier height, i.e. when$\mu \le V_0$.

\vspace{0.5cm}

\section{Two-mode model with LHY term}\label{dimer}
Below we will study symmetric case $E_1=E_2$. Introducing variables:
\begin{eqnarray*}
u=\rho_1 e^{i\theta_1}, \ v =\rho_2 e^{i\theta_2}, \ Z=\frac{N_1 -N_2}{N}, \\ \theta = \theta_2 -\theta_1, \ N_i = \rho_i^2, \ N=N_1 + N_2,
\end{eqnarray*}
where $Z$ is the atomic populations imbalance, $\theta$ is the relative phase, $N$ is the total number of atoms, indices \{$1,2$\} relate to the corresponding trap potential wells and substituting these variables into Eqs.~(\ref{dimer_eq1}) and (\ref{dimer_eq2}), we obtain a set of equations:
\begin{equation} \label{eqz}
Z_t = -2K\sqrt{1-Z^2}\sin(\theta),
\end{equation}
\begin{equation} \label{eqtheta}
    \theta_t = \bar{\gamma} N Z  + \frac{2K Z}{\sqrt{1-Z^2}}\cos(\theta) + \bar{g} \sqrt{\frac{N}{2}}[(1-Z)^{1/2} - (1+Z)^{1/2}].
\end{equation}
The set of equations has the Hamiltonian:
\begin{equation}
H = \frac{\bar{\gamma} N}{2}Z^2  -2K\sqrt{1-Z^2} \cos(\theta) -\bar{g} \frac{\sqrt{2N}}{3}[(1-Z)^{3/2}+ (1+Z)^{3/2}].
\end{equation}
The hamiltonian equations for the canonical variables $Z,\theta$ are:
$$
Z_t = -\frac{\partial H}{\partial\theta}, \ \theta_t =\frac{\partial H}{\partial z}.
$$

Stationary states are defined by  fixed points of the
set ((\ref{eqz}), (\ref{eqtheta})): $Z_t =\theta_t =0$.
The symmetric states are:
$$\theta_c =2 n\pi.$$
The ground state energy is:
$$
E_0=-2K -\frac{2g}{3}\sqrt{2N},
$$
and the asymmetric states with higher energy are:
$$\theta_c =(2 n+1)\pi.$$
with
$$
E_{ex}=2K - \frac{2g}{3}\sqrt{2N}.
$$
Then the critical population imbalance can be find from the equation:
$$
\bar{\gamma} N Z_c = \frac{2K Z_c}{\sqrt{1-Z^2}}(-1)^n - \bar{g} \sqrt{\frac{N}{2}}[(1-Z_c)^{1/2} -(1+Z_c)^{1/2}].
$$

\subsection{Josephson oscillations}\label{JO}
Now we consider Josephson oscillations near $\langle Z \rangle =0$. In the case of small oscillations, the two-mode theory works very well, and   the expressions for the Josephson frequency can be obtained from Eqs. (\ref{eqz}) and (\ref{eqtheta}).

i) The zero-phase mode($\theta =0$).
Supposing the amplitude of oscillations of $Z$ and $\theta$ to be small in Eqs (\ref{eqz}) and (\ref{eqtheta}) one can find the frequency of Josephson oscillations near $\langle Z \rangle =0$.
\begin{equation} \label{zerophase}
\omega_J = \sqrt{2K(2K + \bar{\gamma} N - \bar{g}\sqrt{N/2})}.
\end{equation}
In dimensional variables:
\begin{equation}
\hbar\omega_J = \sqrt{(2\bar{K})^2+2\bar{K}\alpha N- 2\bar{K}\beta \sqrt{N/2} }.
\end{equation}
Here, $\alpha=\delta g \int \phi_1^4 dx \ $, $ \ \beta=\frac{\pi \hbar}{\sqrt{2m}} \int \phi_1^3$,
$ \ \bar{K} = -\left[\frac{\hbar^2}{2m} \int \phi_{1,x}\phi_{2,x}dx + \int \phi_1 V(x) \phi_2 dx \right]$.

In the case of the LHY fluid when $\bar{\gamma}=0$, the frequency of the Josephson oscillations (JO) becomes less than the Rabi frequency. It can be evidence that the system is in the LHY fluid regime. Note that in the quasi-one-dimensional regime, the JO frequency is higher, than the Rabi oscillations frequency~\cite{AGS,Salas}. Measuring the oscillation frequency then allows us to find the contribution of the quantum fluctuations.

ii) $\pi$-phase mode($\theta =\pi$).
The frequency of the Josephson oscillation is:
\begin{equation} \label{piphase}
\omega_J = \sqrt{2K(2K - \bar{\gamma} N + \bar{g}\sqrt{N/2})}.
\end{equation}
In dimensional variables:
\begin{equation}
\hbar\omega_J = \sqrt{(2\bar{K})^2-2\bar{K}\alpha N + 2\bar{K}\beta \sqrt{N/2} }.
\end{equation}

\subsection{Self-trapping regime}\label{ST}
Self-trapping regime means that in each well the atoms are localized,
i.e. $\langle Z \rangle \neq 0$. ST mode occurs depending on the values of $Z(0)$ and $\gamma$  We can determine the critical values of $Z_c$ and $\theta_c$ using the Hamiltonian.

The transition condition to the self-trapping regime is determined as follows.
Let us rewrite the Hamiltonian by introducing new parameters:
\begin{equation} \label{Hamilton}
\bar{H}=\frac{\Lambda Z^2}{2}-\sqrt{1-Z^2} cos(\theta) -\delta \left[(1-Z)^{3/2}+(1+Z)^{3/2}\right].
\end{equation}
Here, $\bar{H}=H/2K, \ \Lambda=\bar{\gamma}N/(2K), \  \delta=\bar{g}\sqrt{2N}/(6K)$.

Fixing $Z(0),\theta(0)$, we obtain the condition for the self-trapping:
$$
\bar{H} > -1 - 2\delta
$$
Then we can obtain a critical value of the parameter $\Lambda$ when
the self-trapping occurs:
\begin{equation} \label{condition}
\Lambda_c=\frac{2}{Z(0)^2}\left[-1+\sqrt{1-Z(0)^2}cos(\theta)-\delta (2-(1-Z)^{3/2}-(1+Z)^{3/2})\right].
\end{equation}
\begin{figure}[htbp]
\includegraphics[width=0.3\textwidth]{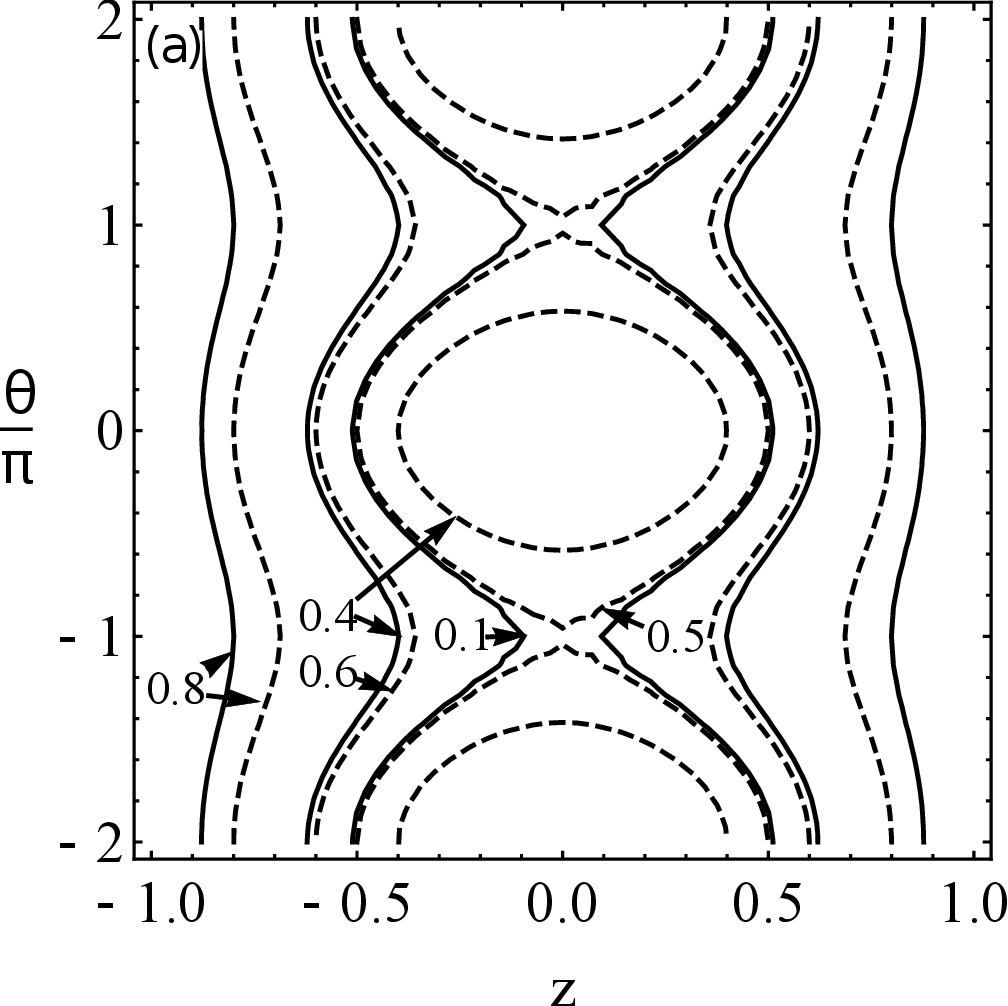}\quad
\includegraphics[width=0.3\textwidth]{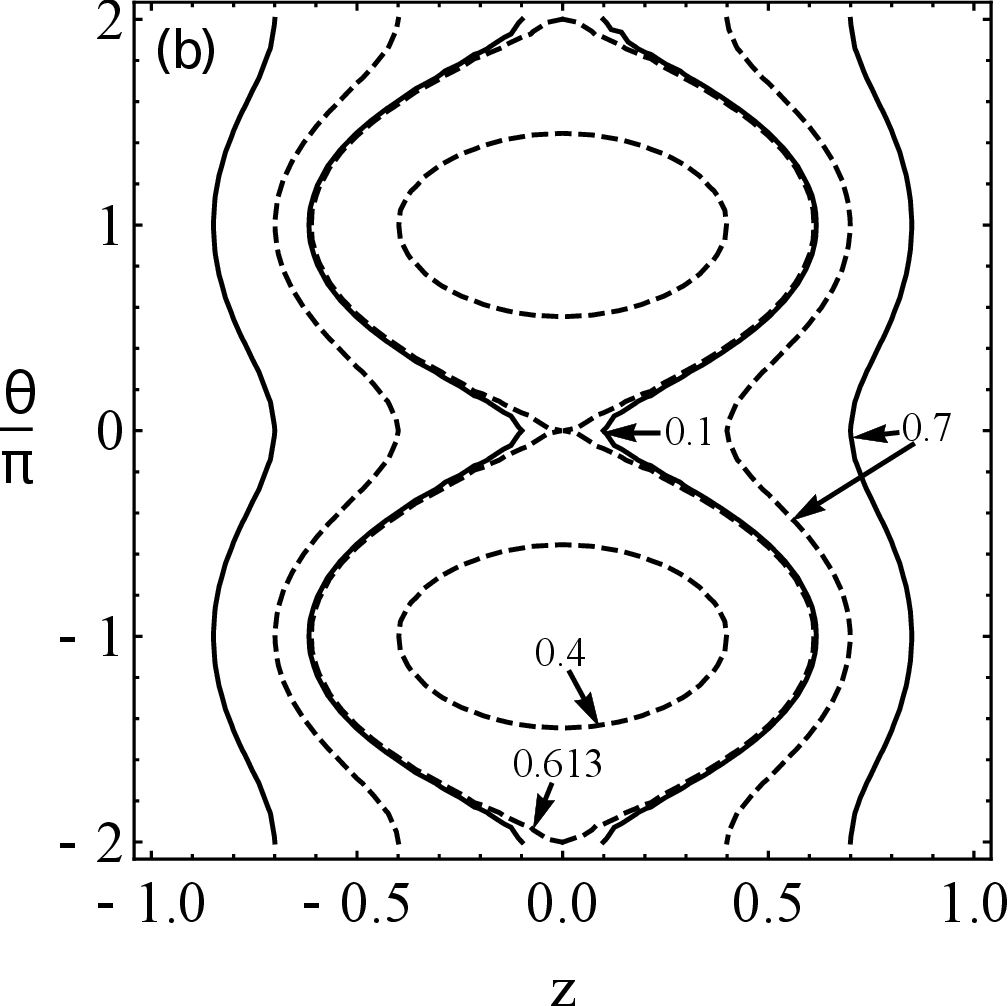}\quad
\includegraphics[width=0.3\textwidth]{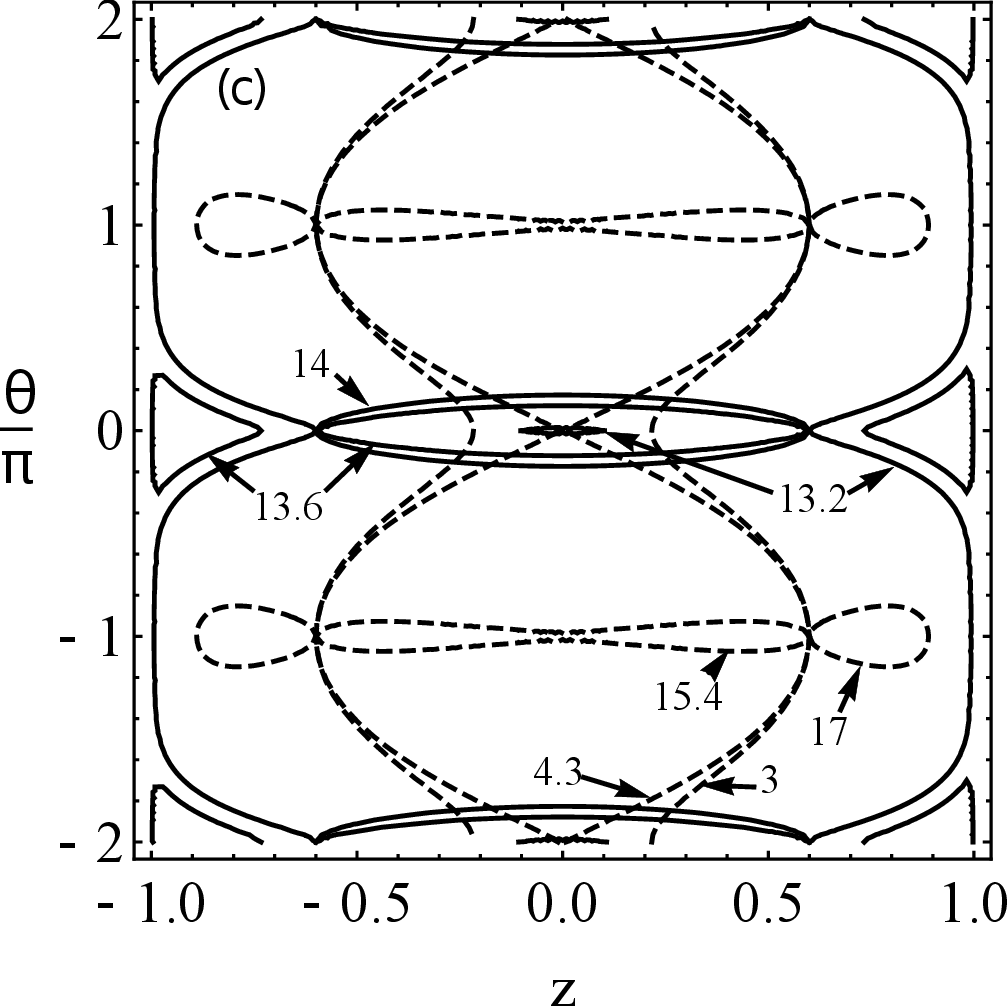}\quad
\caption{First two phase portraits are shown for the parameters $\Lambda=0$, $\delta=9.6$
(LHY fluid)(a) and $\Lambda=4.81$, $\delta=9.3$(b) in the different initial
values of the relative imbalance $Z(0)$. Third phase portrait is presented for
different values of $\Lambda$ when $\delta=9.3$ and $Z(0)=0.6$ (c). In all cases,
the dashed line is for $\pi$-phase mode and the solid line is for zero-phase mode.}\label{fig2}
\end{figure}
In Fig.~\ref{fig2}(a),  the phase portraits is plotted using
Eq.~(\ref{Hamilton}) for the particular case of so-called
LHY fluid with $\gamma=0$  at different initial values of the
relative imbalance $Z(0)$. The LHY fluid corresponds to the case when the
residual mean field is equal to zero $\Lambda = 0$, so the nonlinearity in
the GP equation is given only by quantum fluctuations~\cite{Jor,Skov}.
From this graph we can determine the critical value of $Z(0)$, $Z(0)_c\approx0.504$
for $\pi$-phase mode. This critical value satisfies Eq.~(\ref{condition}).
In the zero-phase mode, there is a localization of the relative imbalance $Z(t)$
for all initial values of $Z(t)$. The relative phase is not localized even in
the self-trapping regime in both zero- and $\pi$-phase modes.

In Fig.~\ref{fig2}(b), the phase portrait is presented for the
parameters $\Lambda=4.81$, $\delta=9,3$ in different initial values
of $Z(t)$. This phase portrait is similar to the first one, only here the
critical value for Z(0) is slightly larger than the first one, $Z(0)_c=0.613$.
This critical value satisfies Eq.~(\ref{condition}).

In Fig.~\ref{fig2}(c), the phase portrait is presented for different values
of $\Lambda$ when $\delta=9.3$ and $Z(0)=0.6$. Both phase modes have
Josephson oscillations and ST mode. We can determine the critical values using
Eq.~(\ref{condition}),
$\Lambda_c\approx 13.2$ for the zero-phase mode and $\Lambda_c\approx 4.3$ for the $\pi$-phase mode.
Of course, another critical value can be seen from the graph in the $\pi$-phase
mode $\Lambda=15.4$, but this point is unstable based on Eq.~(\ref{piphase}).

\begin{figure}[htbp]
\centering
\includegraphics[width=0.9\textwidth]{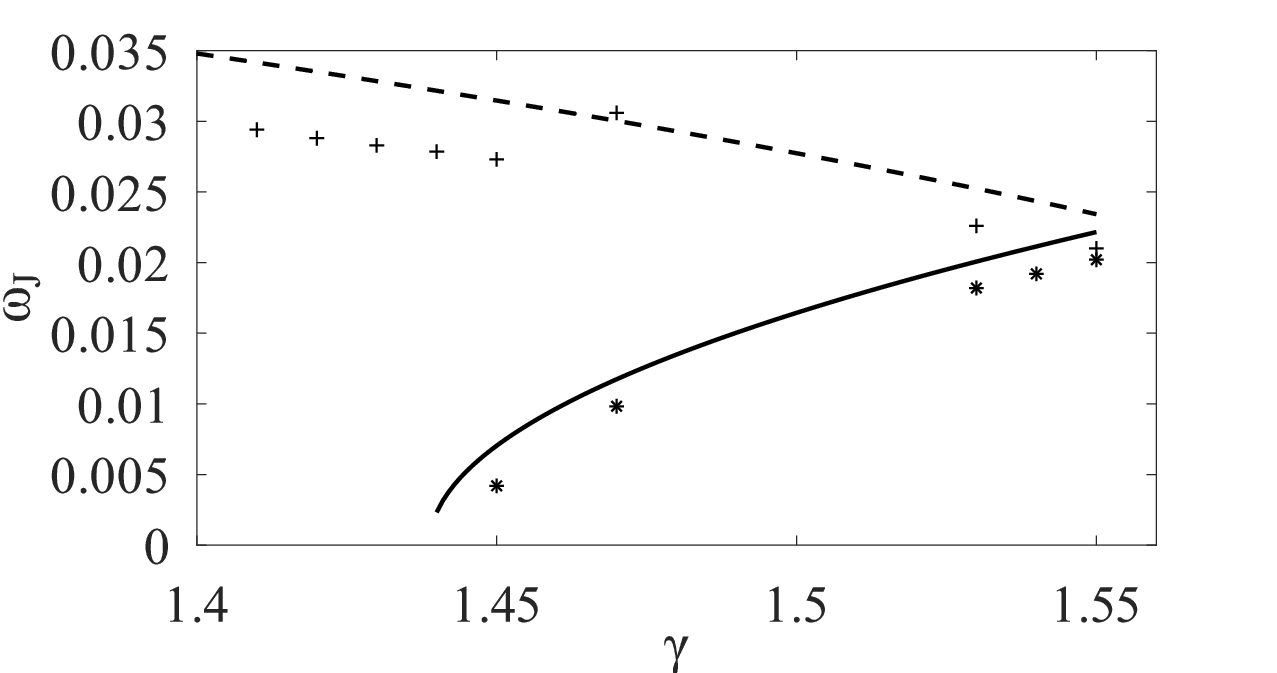}\quad
\caption{Dependencies of the frequency of the Josephson oscillations on the residual mean-field parameter $\gamma$. The solid and dashed lines stand
for the zero-phase, Eq.~(\ref{zerophase}), and $\pi$-phase modes, Eq.~(\ref{piphase}), respectivelly.
The asterisk and symbols $+$ are calculated using numerical
simulations for the zero-phase and $\pi$-phase modes, respectively.}
\label{fig3}
\end{figure}

The dependence of the frequency of the Josephson oscillations on the residual
mean-field interaction strength is presented in Fig.~\ref{fig3}.
The figure shows a good agreement of the two-mode model predictions with full simulations for $\gamma$ around $1.52 -1.55$.
According to Eq.~(\ref{zerophase}), the Josephson frequency becomes complex at $\gamma \leq 1.43$. This can also be seen from the lower curve in the graph. Therefore, when $\gamma \leq 1.43$,
the Josephson oscillation does not exist, where only the ST regime exists. If we change the value of $\gamma$ to $\Lambda$, we can compare this result with the phase portrait results in Fig.~\ref{fig2}, for $\gamma=1.43$, $\bar{\gamma} \approx 1.43 \times 0.19364$, $\bar{g} \approx 0.42625$, $K \approx 0.0114$, $\Lambda=\bar{\gamma}N/(2K) \approx 9.6$. It can be seen from Fig \ref{fig2}a,b, ST regime exists for all values of $Z(0)$, with $\Lambda \leq 9.6$ for zero-phase mode.
The upper curve in Fig.~\ref{fig3} corresponds to the $\pi$-phase mode. In this mode, the Josephson frequency is not complex, so, both, the regime of the Josephson oscillations and the ST regime can exist. This result corresponds to the result in Fig.~\ref{fig2}.
\begin{figure}[htbp]\label{fig4}
\centering
\includegraphics[width=0.3\textwidth]{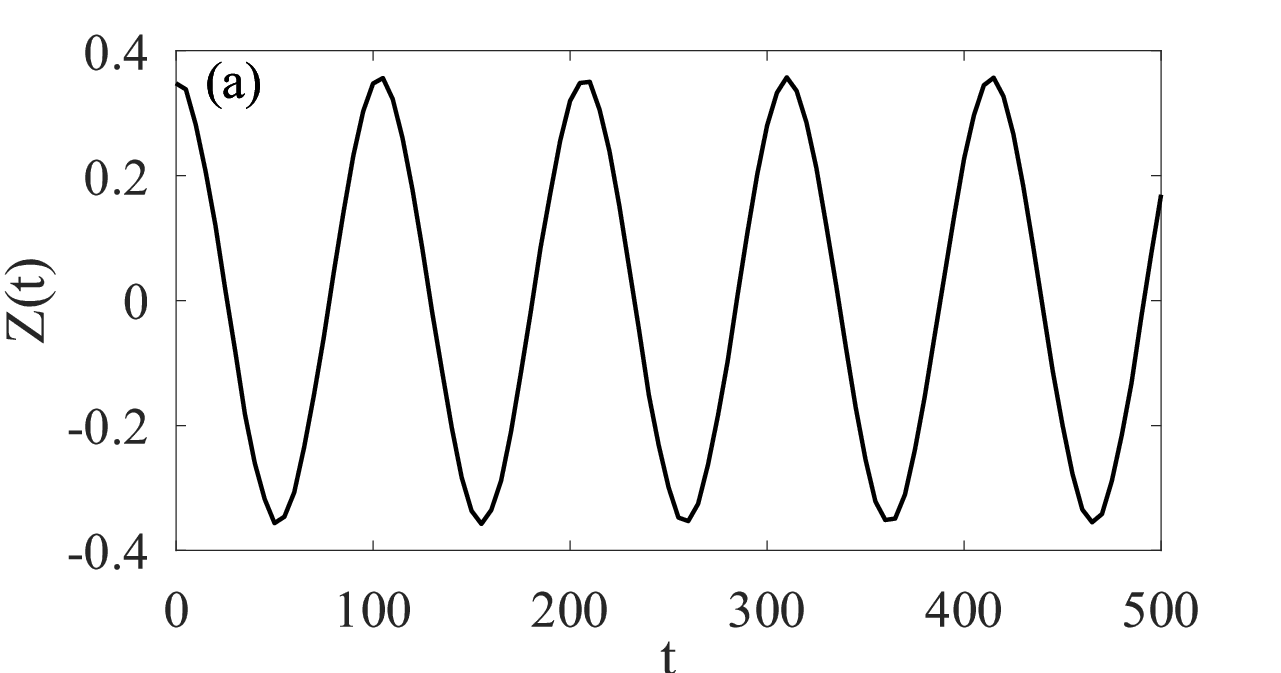}\quad
\includegraphics[width=0.3\textwidth]{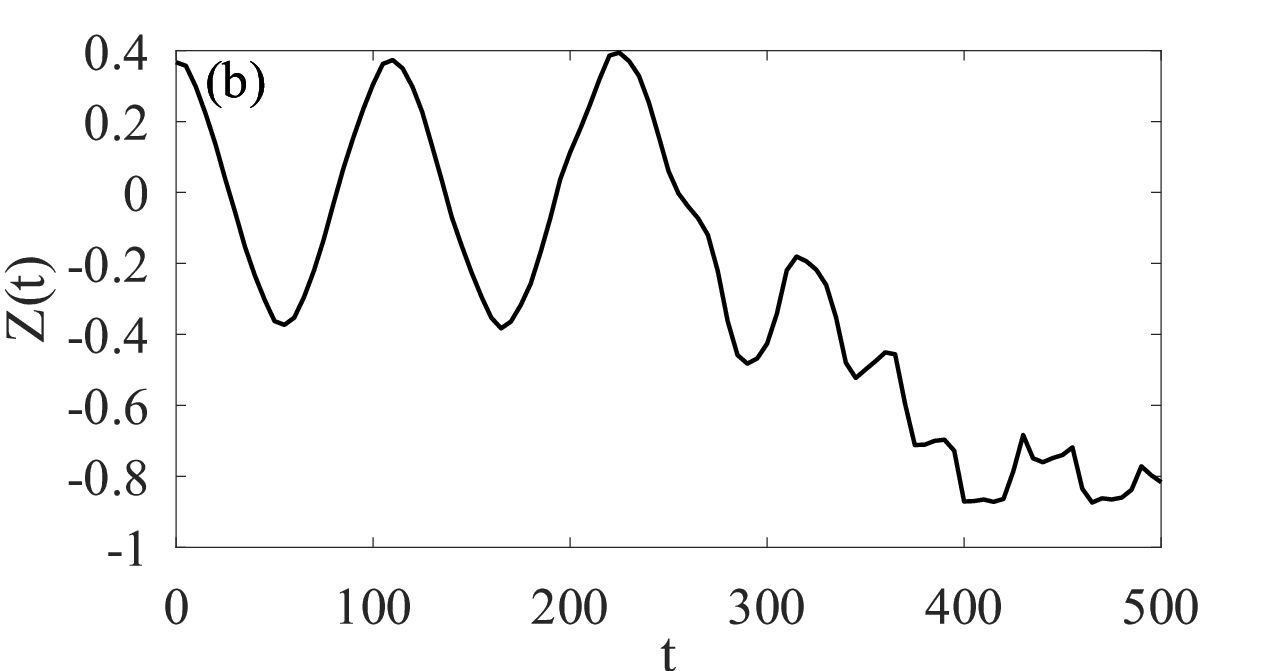}\quad
\includegraphics[width=0.3\textwidth]{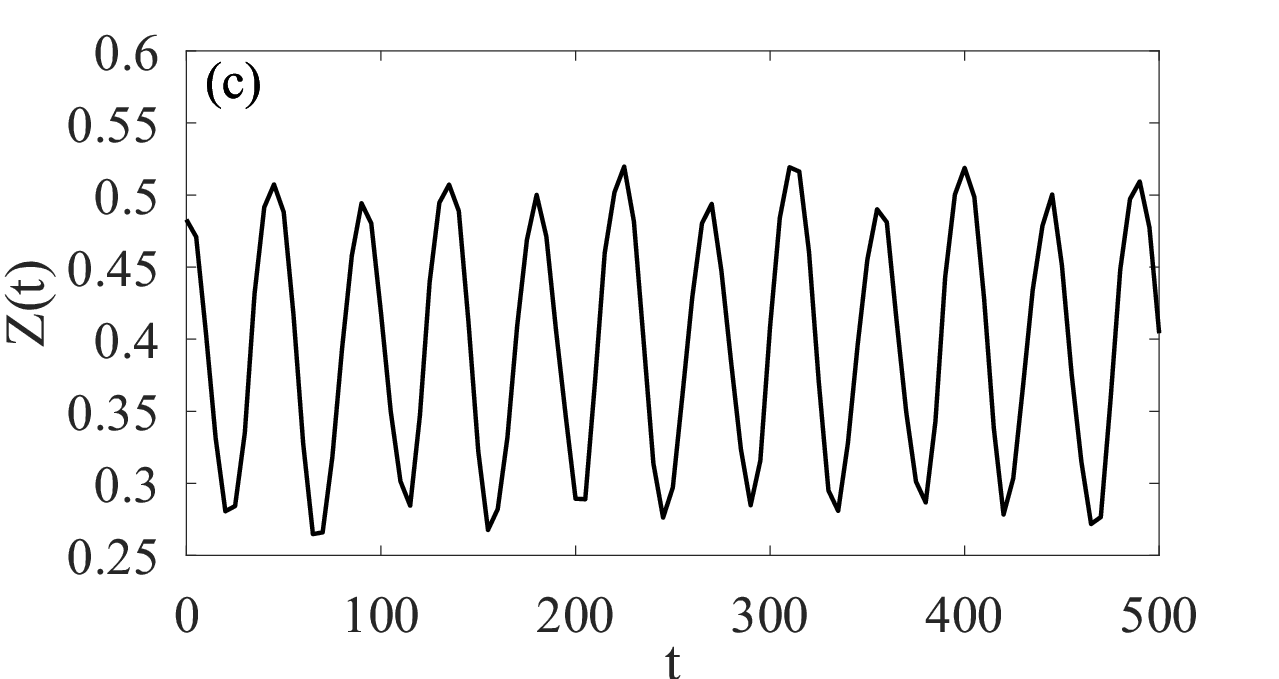}\quad
\caption{The dynamics of the relative imbalance ($Z(t)$) is presented for
LHY fluid case ($\gamma=0$). Here, $Z(0)=0.36$ (a), $Z(0)=0.38$ (b),
$Z(0)=0.5$(c). For all cases $V_0=1$, $\Omega=0.1$, $l=0.5$.}
\end{figure}
The graphs Figs~\ref{fig4}a,b,c  depict the transition of the dynamics of imbalance Z(t) from
the Josephson oscillation regime to the ST regime for the LHY fluid ($\gamma=0$). We can determine the critical value of $Z(0)$, $Z(0)=Z_c=0.38$ in these graphs.
From the figures, it can be seen that Josephson oscillations appear, when the condition $Z(0)<Z_c$ is met. When the condition $Z(0)>Z_c$, the dynamics of $Z(t)$ goes into the ST regime.
\begin{figure}[htbp]
\includegraphics[width=0.3\textwidth]{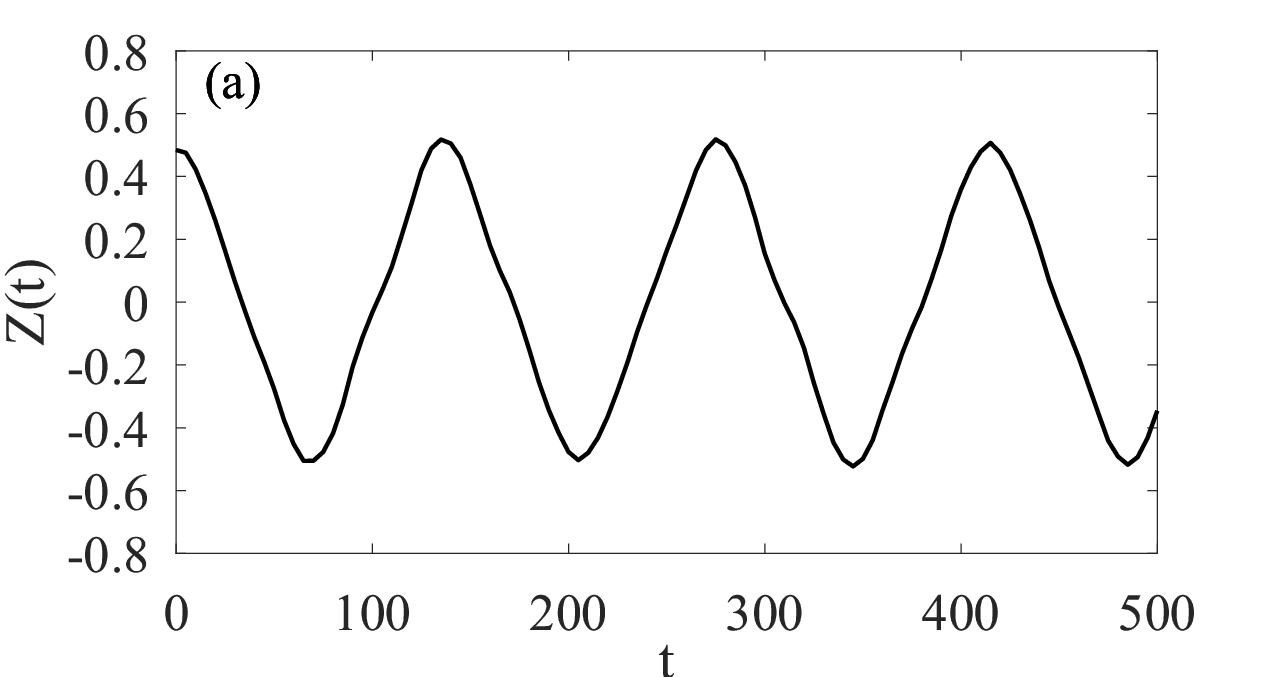}\quad
\includegraphics[width=0.3\textwidth]{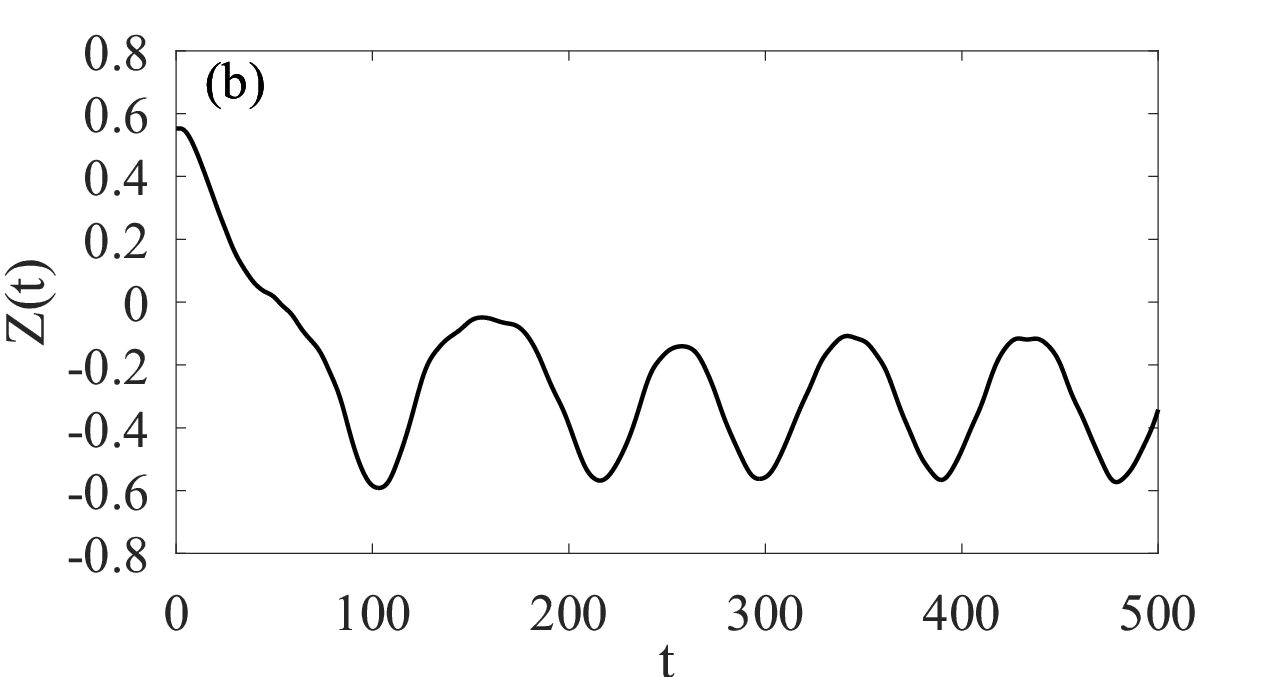}\quad
\includegraphics[width=0.3\textwidth]{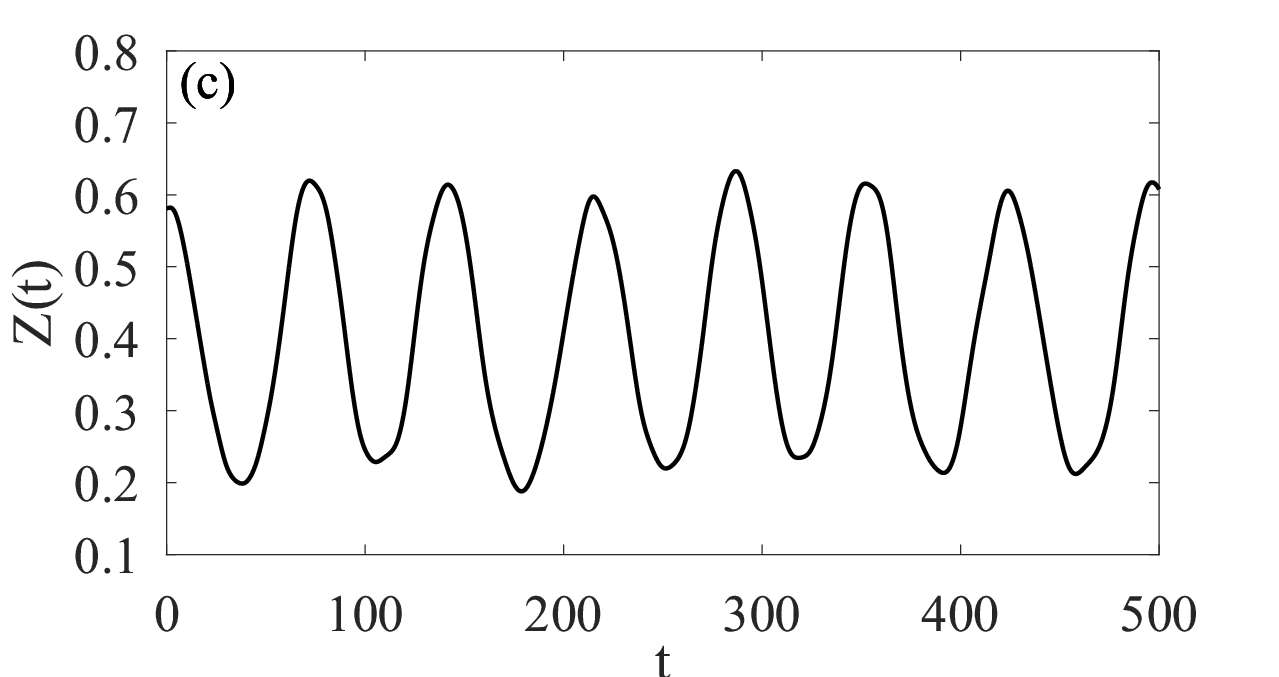}\quad
\caption{The dynamics of the relative imbalance ($Z(t)$) are presented for $\gamma=0.5$ $(\Lambda \approx 4.81)$.
Here, $Z(0)=0.5$ (a), $Z(0)=0.57$ (b), $Z(0)=0.6$(c). For all cases $V_0=1$,
$\Omega=0.1$, $l=0.5$.} \label{fig5}
\end{figure}
The graphs Figs~\ref{fig5}a,b,c depict the transition
of the dynamics of imbalance Z(t) from the Josephson oscillation regime to
the ST one for $\gamma=0.5$). We can determine the critical value
of $Z(0)$, viz. $Z(0)=Z_c=0.57$ in these graphs. The Josephson oscillations appear, when the condition $Z(0)<Z_c$ is met, and the dynamics of $Z(t)$ goes into the ST regime. The two-mode model predicts that the threshold value of the $\bar{\gamma}$ is increased in the one-dimensional geometry when the LHY correction is taken into account, at the same time in the quasi-one-dimensional case, it is reduced, in comparison with the mean-field value~\cite{Salas,AGS}. The full numerical simulations confirm this prediction. But theoretical and numerical values agreed only quantitatively, due to well-known limitations of the two-mode model~\cite{AB06}.

\section{Modulated in time tunneling}
\label{sec:resonance}
Let us study the Josephson oscillations when the barrier is periodically modulated in time:
\begin{equation}
    V(x)=\frac{\Omega^2}{2}x^2+(V_0+V_1 sin(\omega t)) sech^2\left(\frac{x}{l}\right).
\end{equation}
Such modulation leads to the oscillating in time of the macroscopic tunneling coefficient $K$:
\begin{equation}
K(t) = K_0 + K_1\sin(\omega t),
\end{equation}
where
$$
K_0=-\left[\frac{1}{2} \int dx \phi_{1,x} \phi_{2,x} + \int dx \phi_1 V_0 \phi_2 \right], \  K_1 =-\left[\int dx \phi_1 V_1 sech^2(x/l)\phi_2 \right]
$$.
This procedure represents one of the ways to control the macroscopic quantum
tunneling of BEC in the double-well trap~\cite{AK2000,AK,Melburn,Barboza}.

Considering small oscillations of $Z,\theta $ near zero values, we obtain
the equation for the populations imbalance:
\begin{equation}
\delta Z_{tt}+ \omega_0^2(1 + h\sin(\omega t))\delta Z -\frac{K_1\omega}{K_0}\cos(\omega t)\delta Z_t =0,
\end{equation}
where
$$
\omega_0^2 = 2K_0\left(2K_0 + \gamma N + \sqrt{\frac{N}{2}}\right), \ h = \frac{K_1}{K_0}\left(1+\frac{4K_0^2}{\omega_0^2}\right).
$$
The standard analysis \cite{LL}, shows that the parametric resonance in
the oscillations of the imbalance exists at $$\omega = 2\omega_0.$$
The gain at the parametric resonance $\lambda$ is:
\begin{equation}
\lambda =\sqrt{\beta - \frac{\delta^2}{4}}, \ \ \beta = \left(\frac{K_1}{8K_0}- h\right)\frac{\omega_0^2}{4}.
\end{equation}
Note, that when the nonlinearities are equal to zero, in the case of Rabi oscillations,
the gain is equal to zero and parametric resonance (PR) is absent. It can be seen also from
the initial dimer equations Eq.~(\ref{dimer_eq2}) in the linear limit, since in this case
the modulations of $K(t)$ can be removed from the equations by
redifinition of the time as $\tau = \int K(t)dt$.
One of the important consequences is that using the PR for the LHY fluid
(with $\gamma =0$~\cite{Jor}) we can measure the strength of quantum fluctuations.

\begin{figure}[htbp]
\includegraphics[width=0.45\textwidth]{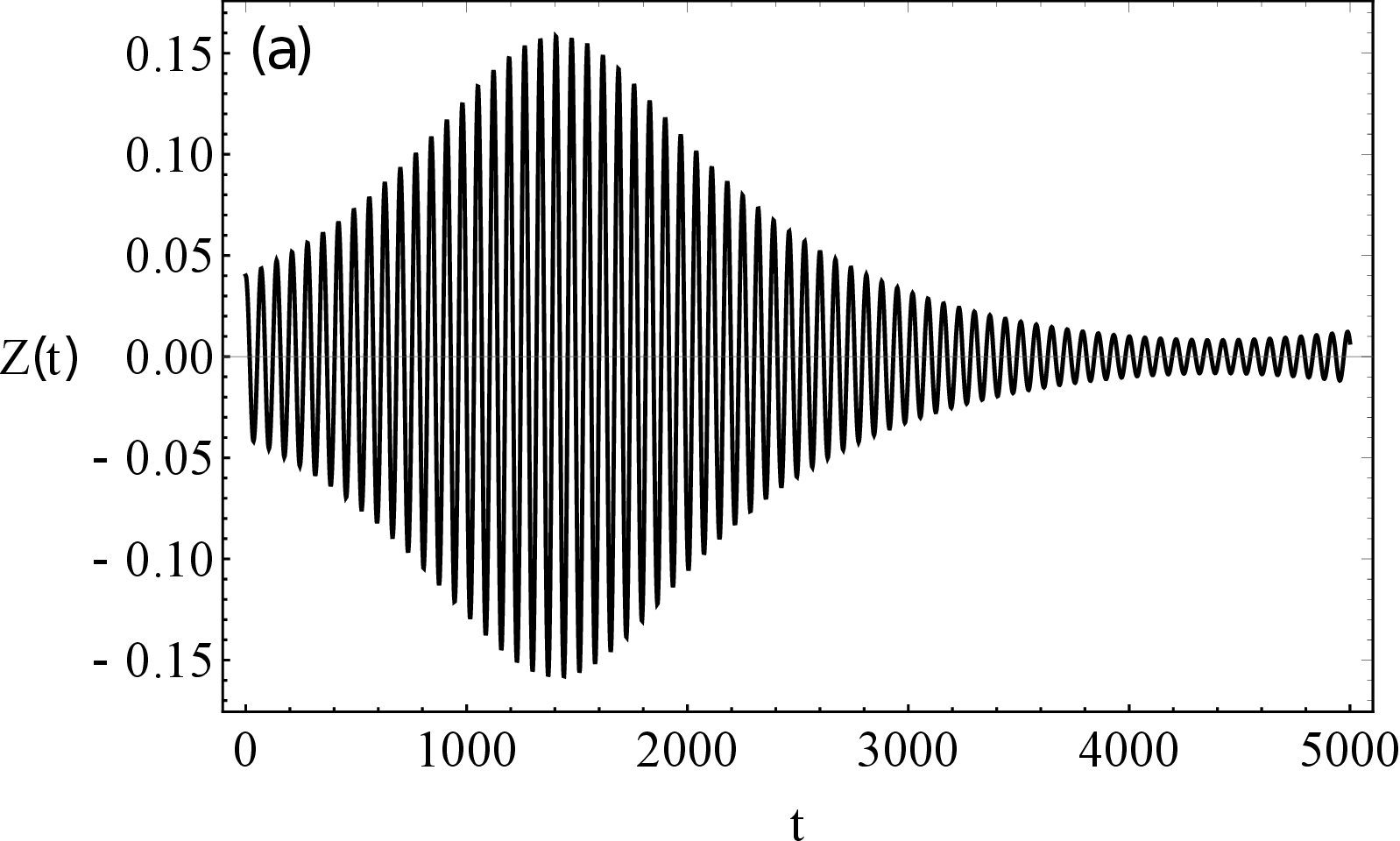}\quad
\includegraphics[width=0.45\textwidth]{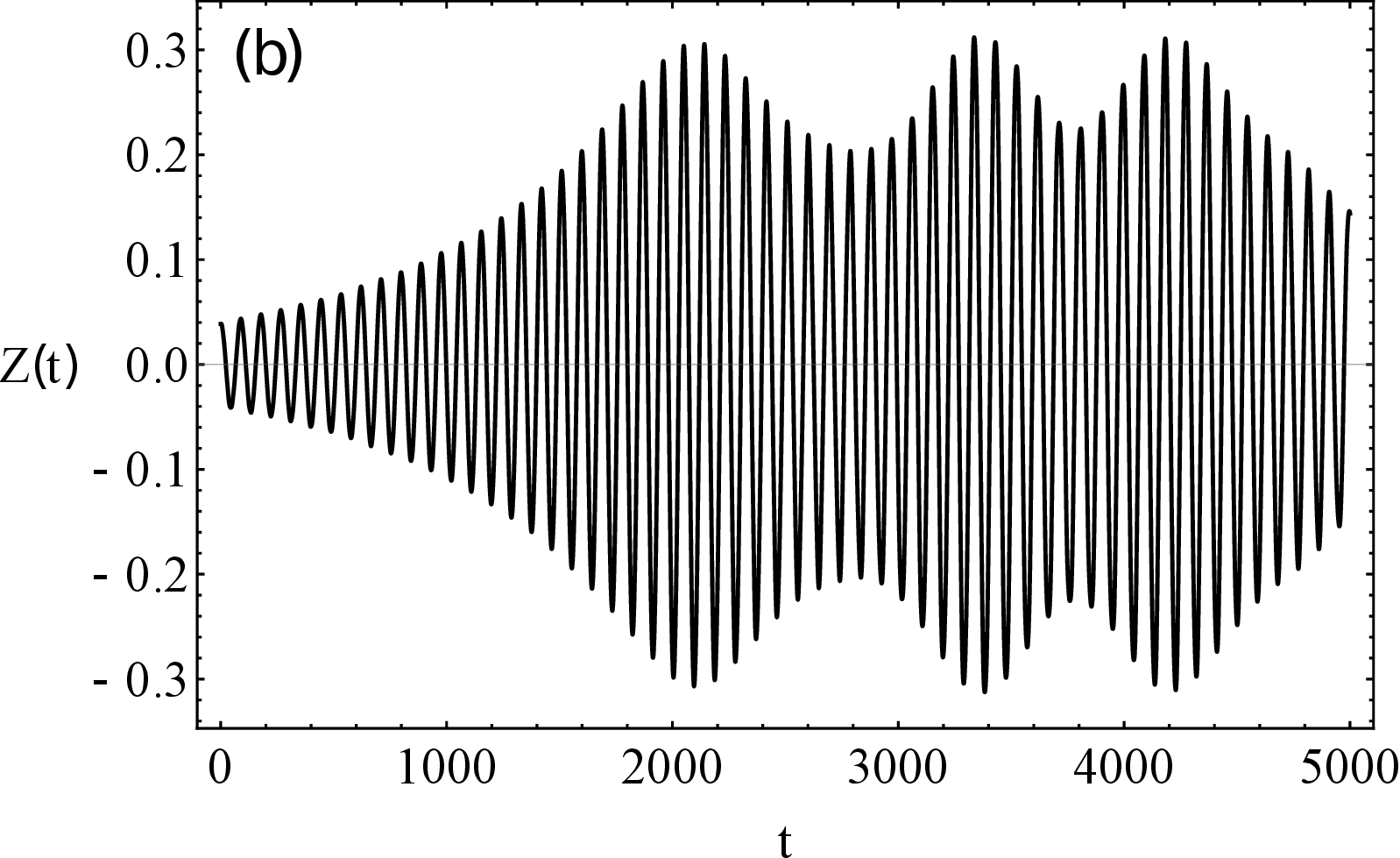}\quad
\caption{The parametric resonances for $\Omega=0.1$,$V_0=1$, $l=0.5$, $V_1=V_0/20$, $\omega=2\omega_J$, are shown at $\gamma=0$. These results are obtained using the two-mode theory (a) and the numerical simulations method (b).} \label{fig6}
\end{figure}

\begin{figure}[htbp]
\includegraphics[width=0.45\textwidth]{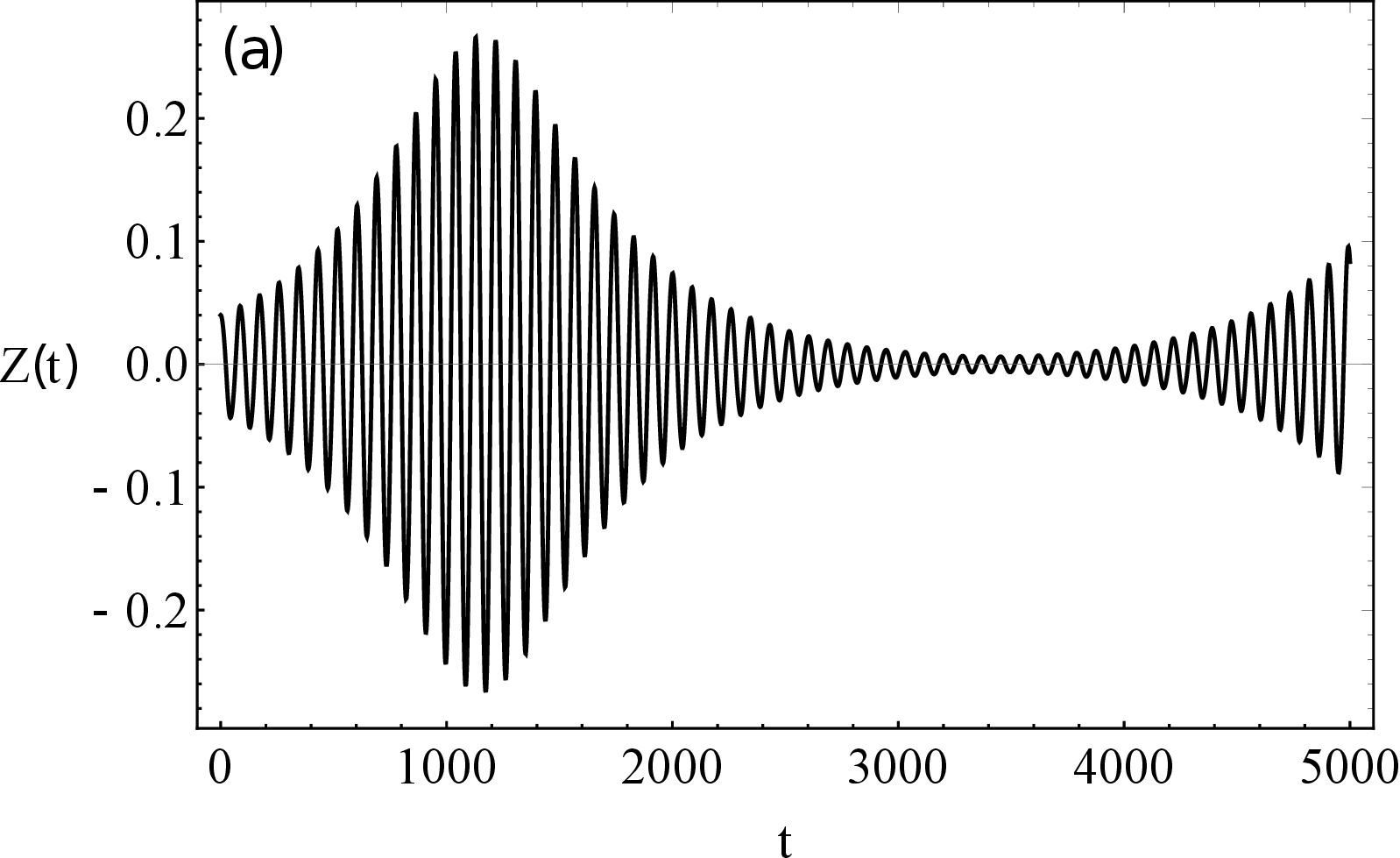}\quad
\includegraphics[width=0.45\textwidth]{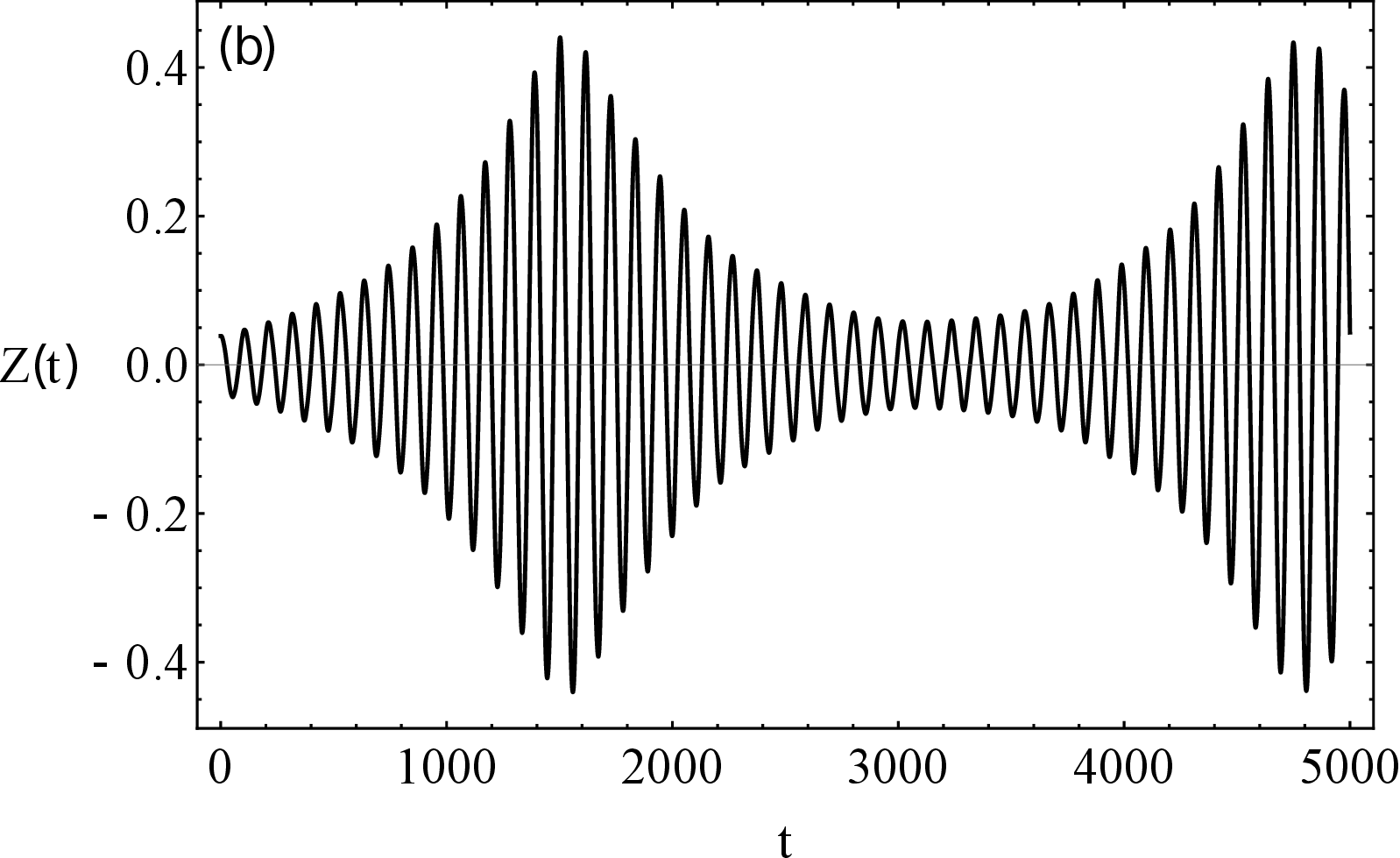}\quad
\caption{The figure depicts parametric resonances for $\Omega=0.1$,$V_0=1$, $l=0.5$, $V_1=V_0/10$, $\omega=2\omega_J$, at $\gamma=0.5$. Results are obtained using the two-mode theory (a),
and the numerical simulations method (b).} \label{fig7}
\end{figure}

In Figs ~\ref{fig6} and~\ref{fig7} the results of numerical simulations obtained
from the two-mode theory and full GP equation with double-well potential are
presented. The gain at the resonance is agreed well, while the long time dynamics is described qualitatively. it is connected with the approximate character of the two-mode
model, neglecting the nonlinear corrections to the tunneling coefficients $K$.

\section{Conclusion}
We have investigated the nonlinear dynamics of BEC loaded in a double-well potential.
The beyond mean-field corrections to the energy in the form of the Lee-Huang-Yang term are
taken into account. We derive a two-mode model to describe analytically the dynamics of BEC.
Using this model we obtain the expressions for the frequencies of the Josephson oscillations
when the contribution of the quantum fluctuations are taken into account. The threshold value of the residual mean field nonlinearity, when switching from the MQT regime to the
ST regime occurs, has been found.  The results applied for the description of the dynamics of
the quasi-one-dimensional LHY superfluid in the double-well potential. The parametric resonance in the Josephson oscillations when the barrier is periodically modulated, is studied. The analytical predictions are compared with the full numerical simulations of the modified GP equation,
showing good agreement.

\section{Acknowledgments}
We thank B. B. Baizakov and  E. N. Tsoy for fruitful discussions. This work has been implemented within the research topic of the Laboratory of Theoretical Physics of Physical-Technical Institute of Uzbekistan Academy of Sciences, funded by the State Budget of Uzbekistan (Grant No. 2024 year award).
F.Kh.A. acknowledges the support of the Ministry of Higher Education, Science and Innovation of the Republic of Uzbekistan (Grant No. ALM-2023-1006-2528).
 \vspace{1cm}
\section{Author contributions}
All the authors contributed equally to this work.
All the authors have read and approved the final
manuscript.

\end{document}